# Beneath and beyond frustrated total reflection: a practical demonstration

Carolina Filgueira-Rama, Alejandro Doval, Yago Arosa and Raúl de la Fuente

Grupo de Nanomateriais, Fotónica e Materia Branda, Departamento de Física Aplicada, Universidade de Santiago de Compostela, E-15782, Santiago de Compostela, Spain.

Abstract

Frustrated total internal reflection is analyzed from an unusual point of view Unlike most similar works, incident angles are used here as the scanning variable, instead of the tunneled film thickness. The theoretical framework is presented defining our own normalized variables, better adapted to the problem at hand. A straightforward and captivating experiment, appropriate for undergraduate classroom demonstrations is also presented. The experiment involves measuring the reflection and transmission of light through a pair of prisms separated by an air or water layer, and the results are in fair agreement with theory.

## 1. Introduction

Total internal reflection (TIR) at an optical interface between two transparent mediums is a remarkable phenomenon. When a light beam refracts from a transparent medium towards another with a lower optical density at a sufficiently oblique angle, the interface behaves like a perfect mirror. This allows us, for example, to see reflected objects inside a water tank when observing the surface from below. It is also the reason why some polyhedral gemstones show vibrant colors. TIR has many useful applications for our daily life. For instance, optical fibers, commonly used in communications or in an endoscope, rely on total reflection to guide light [1]. It is also key in erecting prisms which are employed to invert images inside binoculars with minimal light loss, in periscopes which allow us to look at raised objects from a concealed position, or in rain sensors for automatic windshield wipers in cars [2].

TIR occurs when the incident angle $\theta$ is greater than the critical angle $\theta_{cr}$, which is a function of the refractive indices of the two mediums $(n_1, n_2)$: $\theta_{cr} = \arcsin(n_2/n_1)$, being $n_2 < n_1$. However, even if the condition $\theta > \theta_{cr}$ is fulfilled, the electric field of the incoming wave penetrates the second medium in the form of an evanescent wave. The penetration depth is in the order of a wavelength and can be observed by introducing an optically denser third medium $(n_3 > n_2)$ close after the first interface, so that the second medium becomes a thin layer. If the width of the low-index layer $d$ is small enough ($d \lesssim \lambda$), the evanescent wave can couple to the new medium, carrying energy that will propagate through it. If this occurs, the total internal reflection is said to be "frustrated". From now on we will refer to this phenomenon as FTIR (Frustrated Total Internal Reflection) as it is usually done in literature. We would like to emphasize its interest, since, quoting [3], FTIR "demonstrates the wave nature of light in a unique fashion".

Several valuable applications have been found for FTIR as well. Digital fingerprint sensors, surface microscopy, beam splitters, light couplers or optical filters are some examples of its great practical use [4-7]. In addition to this technological interest, FTIR has also been widely studied because of its undeniable value from an educational perspective. The effect has been presented as analogue to quantum tunneling by many authors, such as [8 - 10], since it can constitute a simpler and visually direct demonstration for an undergraduate laboratory or classroom lesson [11].

The works on FTIR are typically focused on analyzing the frustration as a function of the thickness $d$ of the low-index layer. That requires being able to precisely control sufficiently small variations in $d$, which means, for visible light, absolute widths of around or less than a micron. Some authors [11 - 13] vary $d$ sequentially, while others [8, 14] consider configurations where $d$ varies along the interface, i.e., by using a linear or a curved wedge. To facilitate the experimental arrangement, the phenomenon has also been studied for longer wavelengths, such as microwaves [15], for which separation between the two high index mediums should be in the range of millimeters and is easily adjustable. In any case, visible light is preferred because of its visual impact. Our idea is to review the FTIR phenomenon from a different point of view using a constant low-index layer width and analyzing FTIR as a function of the incident angle. This means switching from thickness control to angle control, which has been done using a medium-resolution rotating platform (namely 0.1°). This angle shifting method allows us to explore at the same time what happens beneath and beyond the critical incidence angle. When scanning different angles of incidence, the measured transmittance through the three mediums system can be seen to appreciably fluctuate until it suddenly decreases in the FTIR zone, gradually approaching zero for very oblique angles of incidence. If there was no frustration of the evanescent tail, transmittance would immediately drop to zero as soon as the incident angle reaches its critical value.

In this article, we discuss our idea and show some experimental results following the subsequent structure. Firstly, electromagnetic theory concerning FTIR is revisited in section 2, focusing on the expression for transmittance as a function of the incident angle. Section 3 is the core of our manuscript, where various FTIR experiments that clearly demonstrate the phenomenon are presented, together with an analysis of the obtained experimental results is conducted. In section 4, our FTIR approach is compared to quantum tunneling effect, applying the usual analogy between both phenomena to our particular case. Finally, some overall conclusions are presented.

## 2. Theoretical background: modelling Frustrated Total Internal Reflection (FTIR)

FTIR theory in the visible region of the electromagnetic spectrum has been revisited in several scientific papers [8, 12, 13]. Most of them study reflection and refraction of light when it propagates through a three-layer system involving three transparent mediums, being the second medium the one with a lowest refractive index ($n_L$, where the subindex $L$ stands for 'low'). It is common to consider, the symmetrical case, in which the third medium is the same material as the first one. This situation consists in a thin low-index dielectric layer, typically air, separating two equal optically denser dielectric mediums ($n_L < n_H$, where the subindex $H$ stands for 'high', since their refractive indices are higher). The corresponding refractive index profile is represented in Fig. 1, being the x direction normal to both interfaces:

$$n(x) = \begin{cases} n_H & x \leq 0 \\ n_L & 0 < x \leq d \\ n_H & x > d \end{cases} \qquad (1)$$

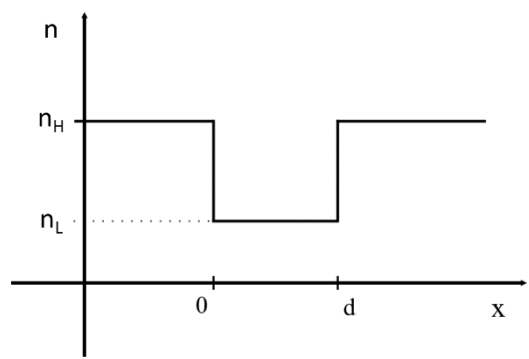

Figure 1. Considered symmetrical refractive index profile with a low dielectric index layer.

The expressions for the electric fields of the waves can be obtained particularizing Helmholtz's equation for the refractive index profile given in Eq. (1). The planar symmetry in the considered problem allows us to reduce Helmholtz's equation to a one-dimensional expression, depending only on the $x$ coordinate. Although a comprehensive study would involve analyzing both transverse electric (TE) and transverse magnetic (TM) waves, we focus solely on the former, as the results are similar in both cases. TE waves are those polarized in such a way that their electric field vibrates perpendicular to the plane of incidence. Taking the $xz$ plane as the plane of incidence, the electric field for TE polarization would be in the y direction. Therefore, one-dimensional Helmholtz's equation for TE waves in our case states:

$$\frac{d^2 E_y(x)}{dx^2} - [\beta^2 - k_0^2 \, n^2(x)] \, E_y(x) = 0 \tag{2}$$

where $k_0 = \omega/c = 2\pi/\lambda$ is the wavenumber of the waves in vacuum (being $\omega$ and $\lambda$ the frequency and wavelength, and $c$ the speed of light in vacuum), and $\beta$ is an invariant that must be conserved at the interface for all waves to fulfil the boundary conditions for electromagnetic waves [16]. This conserved quantity corresponds to the component of the wavevectors tangential to the boundary, and its conservation defines the laws of reflection and refraction. Its value can be written in terms of the incident ($i$), reflected ($r$) or transmitted ($t$) wave at the first interface in our problem as follows:

$$\beta = k_H \sin \theta_i = k_H \sin \theta_r = k_L \sin \theta_t$$

$$k_j = k_0 \, n_j \qquad j = H, L \tag{3}$$

being $k_H$ and $k_L$ the wavenumbers for propagation inside each material and being $\theta_i$, $\theta_r$ and $\theta_t$ the angles of incidence, reflection, and refraction, respectively. Since $\theta_i$ is always a real angle, the value of $\beta$ is restricted to the interval $[0, k_H]$.

Since the solutions for the electric fields are well known (see for example [9, 13]) and seem more difficult to compare to any possible experimental results, we will focus on the expressions for transmittance. In analogy with the notation typically used for optical waveguides [17], it is useful to define the following normalized variables, as a tool to simplify the analysis:

$$b = \frac{\beta^2 - k_H^2}{k_L^2 - k_H^2} = \frac{n_H^2 \cos^2 \theta_i}{n_H^2 - n_L^2} \qquad v = d\sqrt{k_H^2 - k_L^2} = \frac{2\pi d}{\lambda} \sqrt{n_H^2 - n_L^2} \tag{4}$$

The value of $v$ (the so-called normalized frequency in the context of optical waveguides, but which we prefer to call normalized thickness here) depends on the thickness of the low-index layer, the difference between the squared refractive indices of the materials, and the wavelength. Meanwhile, $b$ stands for a normalized propagation constant. Its possible values are restricted by the ones $\beta$ can take, being thus confined to the interval $[0, \, n_H^2/(n_H^2 - n_L^2)]$. While in the previous works that have already been mentioned $v$ was a variable, our technique keeps it constant, being $b$ the interrogation variable. We can also define a normalized independent variable $u$,

$$u = x\sqrt{k_H^2 - k_L^2} = \frac{2\pi x}{\lambda} \sqrt{n_H^2 - n_L^2} \tag{5}$$

obtaining a normalized Helmholtz's equation for TE waves:

$$\frac{d^2 E_y}{du^2} + b\, E_y = 0 \qquad \text{for } u \leq 0 \text{ and for } u > v$$

$$\frac{d^2 E_y}{du^2} + (b-1)\, E_y = 0 \qquad \text{for } 0 < u \leq v \tag{6}$$

The expression for the transmittance for TE waves can be easily obtained using its definition after calculating the electric fields using Eq. (2) or, more directly, Eq. (6). It is usually given in terms of the Fresnel coefficients at both interfaces, as it is done for example in [8, 18]. We must consider two different situations, for incident angles $\theta_i$ below and above the critical angle $\theta_{cr}$. In the first case, $\theta_i < \theta_{cr}$, we know that the invariant $\beta$ will also be below its critical value $\beta_{cr} = k_H \sin\theta_{cr} = k_L$. Therefore, $b$ can take values within the interval $(1,\ n_H^2/(n_H^2 - n_L^2)]$, being $b = 1$ when $\theta_i = \theta_{cr}$ and decreasing for larger angles. The expression for transmittance would be:

$$T = \frac{1}{1 + \sin^2(v\sqrt{b-1})/[4b(b-1)]} \tag{7}$$

This transmittance for small incident angles $\theta_i$ presents oscillations depending on the incident angle, resembling the expression corresponding to a Fabry-Perot interferometer [19, 20] with low reflectance mirrors (the interfaces between the different refractive indices). As we increase the angle of incidence at the first interface and approach its critical value $\theta_{cr}$, the reflectance increases rapidly, and we observe a region of incidence angles where the transmittance oscillates between a maximum value $T = 1$ and progressively lower minimums.

For incident angles exceeding the critical angle ($\theta_i > \theta_{cr}$), the conserved quantity is also over its critical value $\beta > \beta_{cr} = k_c$. This leads to a normalized propagation constant restricted to $b \in [0,1)$, and thus the argument of the sine function becomes imaginary. In this second case, transmittance can be rewritten as:

$$T = \frac{1}{1 + \sinh^2(v\sqrt{1-b})/[4b(1-b)]} \tag{8}$$

Although one might expect the whole amount of light to be totally reflected into the first medium and not to reach the third medium, the transmittance of the system is not zero. The existence of the third medium prevents the amplitude of the evanescent wave from fully decaying, giving rise to the frustration of the total reflection. For incidence angles $\theta_i$ really close to (but below) the critical angle $\theta_{cr}$, the transmittance suddenly falls. Once beyond this angle (for $\theta_i > \theta_{cr}$), it gradually decreases to zero at a rate that depends on the thickness $d$ of the low-index layer and on how different the two refractive indices are (i.e. on $v$). The key is that for $\theta_i < \theta_{cr}$ there are waves propagating through the three mediums, while in the case of $\theta_i > \theta_{cr}$ the transmitted wave at the third medium is attenuated in comparison to the incident wave after tunnelling the intermediate layer as an evanescent wave. For completeness the expression for the transmittance at the critical angle $\theta_i = \theta_{cr}$ (for which $b = 1$) is shown:

$$T = T_{cr} = \frac{1}{1 + (v/2)^2} \tag{9}$$

denoting the critical transmittance as $T_{cr}$.

Eqs. (8) and (9) show an inverse dependence of the transmittance on the normalized layer thickness $v$. Therefore, $v$ stands for some kind of measurement of how easily total reflection can be frustrated (the lower $v$, the easier the frustration). In other words, the transmittance will be higher for thinner layers, closer refractive indices and longer wavelengths.

Fig. 2 shows an example of the transmittance as a function of the incident angles at the different regions, including noticeable FTIR over the critical angle. The graph represents the case of an air gap ($n_L = n_{air}$) of thickness $d = 2\lambda$ between two equal glasses with refractive index $n_H = 1.5$. Using Eqs. (4) and (8), the resulting value for the normalized thickness is $v = 14.05$, and the critical transmittance $T_{cr}$ is nearly 2%, as can be clearly seen in the figure.

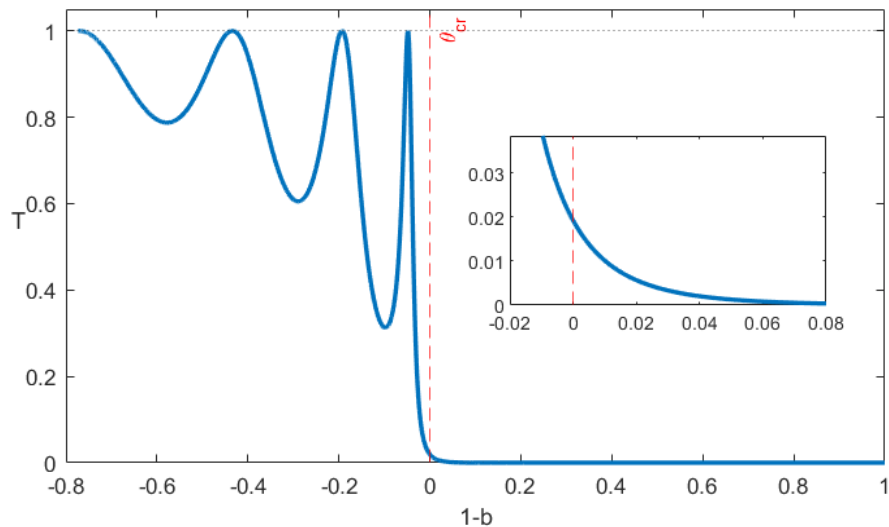

Figure 2. Example of the transmittance as a function of the normalized propagation constant b, including a zoom on the frustrated total internal reflection region for a $d = 2\lambda$ air gap between two $n_H = 1.5$ glasses. The independent variable has been chosen to be $(1 - b)$, since it grows when the incidence angle also does so.

For incident angles below $\theta_{cr}$, the transmittance oscillates as a function of the incident angle, which is clearly predicted in Eq. (7). Its maxima equal $T = 1$, and happen for angles such that $\sin(v\sqrt{b-1}) = 0$ in Eq. (7). This means that constructive interference conditions, correspond to those angles for which half the accumulated phase in a roundtrip within the layer $\delta = v\sqrt{b-1}$ is a multiple of $\pi$. The minima present different values, becoming smaller as $\theta_i$ approaches $\theta_{cr}$, while the width of the peaks decreases, getting both the subsequent maxima and minima closer to each other. The reason for this lies in the reduction of the normalized propagation constant $b$ towards 1, that increases the amplitude $1/b(b-1)$ multiplying the sine function in Eq. (7). On a more physical level, the reflectivity (the surface reflectance) increases gradually towards 1 at $\theta_{cr}$ (see next paragraph for further explanation). Right after the last maximum, the transmitted irradiance suddenly drops towards zero into the total internal reflection region. For incident angles above $\theta_{cr}$, the transmittance decreases gradually from its critical value $T_{cr}$, until it finally vanishes. This fast drop in the transmittance is due to the growth of both $\left[\sinh\left(v\sqrt{(b-1)}\right)/\sqrt{(b-1)}\right]^2$ and, specially, $1/b$. All these remarks can be observed in Fig. 2.

It is also interesting to partially express the transmittance in terms of the reflectivity at the interface, directly related to the corresponding Fresnel coefficient (in this case, the TE Fresnel coefficient) as $R_{sb} = |r_{sb}^2|$, where the subindex $sb$ stands for 'surface boundary'. Let us write the TE Fresnel reflection coefficient at the first interface in terms of the normalized propagation constant $b$:

$$r_{sb} = \frac{\sqrt{b} - \sqrt{b-1}}{\sqrt{b} + \sqrt{b-1}} \tag{10}$$

being $r_{sb} \in \mathbb{R}$ for $b > 1$ (below the critical angle), $r_{sb} = 1$ for $b = 1$ (at critical incidence) and $r_{sb} = e^{i\phi}$ for $b < 1$ (above the critical angle). Therefore, reflectivity increases towards 1 for increasing incidence angles approaching $\theta_{cr}$. Considering this relation in Eq. (10), it is not difficult to write the coefficient of the sine or hyperbolic sine function within the expressions of transmittance in terms of the normalized propagation constant $b$:

$$\frac{1}{4b(b-1)} = \frac{4\,r_{sb}^2}{(1-r_{sb}^2)^2} \tag{11}$$

This shows quite clearly why we say that the contrast between maxima and minima increase for higher reflectivities as $\theta_i \to \theta_{cr}$ (for $\theta_i < \theta_{cr}$). Since the Fresnel coefficient is real in that angular region, its square equals the reflectivity, and the expression in Eq. (11) diverges as $r_{sb}^2 \to 1$ (. However, this does not lead to any physical trouble, since the sine (Eq. 7) or hyperbolic sine (Eq. 8) multiplied by the term in Eq. (11) go to zero at a similar rate yielding the value of the critical transmittance specified in Eq. (9).)

### 3. Experiment and Results

What has been done in the lab is experimentally recreating an example close to that of the theoretical results shown in Fig. 2. Two right-angled H-K9L prisms have been used. H-K9L has similar properties to BK7 but it is more accessible and affordable, being thus suitable for undergraduate classroom demonstrations. The light source was a $2\ mW$ He-Ne laser, with a wavelength $\lambda = 633\ nm$. For such wavelength, the refractive index of the prisms is $n_H = 1.5151$, according to [21].

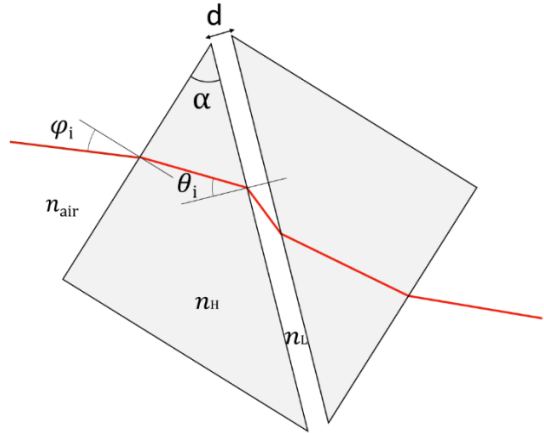

Figure 3. Diagram of the arrangement of the prisms from above. $\varphi_i$ and $\theta_i$ are the external and internal angles of incidence, respectively, and $\alpha$ is the prism apex angle.

In order to appreciate frustrated total internal reflection, the two prisms need to be coupled (see Fig. 3). This coupling was achieved by firmly pressing one prism against the other, facing the two largest faces). Without releasing the pressure on the prisms, they are secured with a clamping arm, as can be seen in Fig. 4. Since the prism surfaces are not perfectly flat, no matter how much pressure is applied, there will be a narrow air gap with variable thickness between them. The refractive index of air is remarkably below the one of the prisms, behaving as the low-index layer, with $n_c \approx 1.00028$ for the chosen $\lambda$ at room temperature [22].

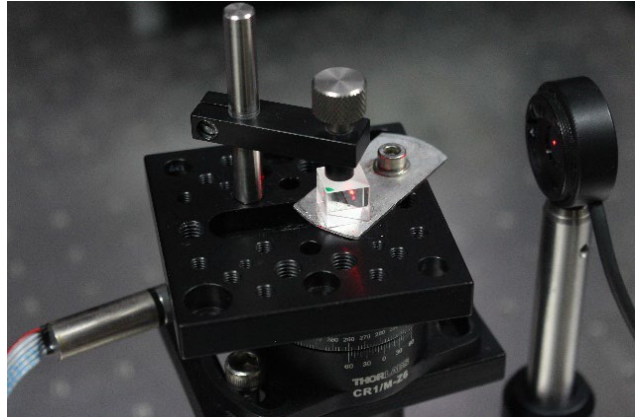

Figure 4. Picture of the experimental arrangement with the two prisms secured by a clamping arm, on top of a rotating platform that allows changing the angle of incidence. On the right side there is a photodiode used as a light detector.

The thinness of the air layer between the prisms can be noticed without any measurements. If the prisms are observed from an appropriate angle, colored interference fringes can be seen, denoting the small differences of the optical path between reflections at each of the two surfaces. This effect is visible in Fig. 5, due to the incidence of white ambient light. All the measurements recorded for the results shown in the subsequent section were taken with all the lights off except for the He-Ne laser source.

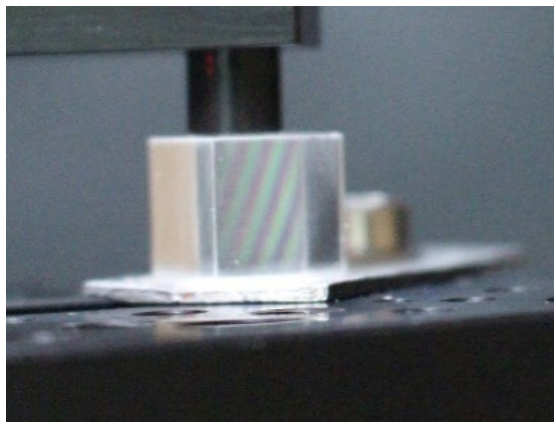

Figure 5. Picture of the experimental setup from such an angle that the interferometric fringes for ambient light reflected in the air gap between prisms are observed.

Aside from air, water has also been used as a low index medium between the prisms. In order to do that, an infiltrating method based on capillarity action is used. A drop of liquid is placed at the base of the contact surface between the prisms, and it is possible to observe how it gradually covers the surface in a few seconds (see Fig. 6). It is easy to notice whether there is water or air between the prisms because of the change in the critical angle.

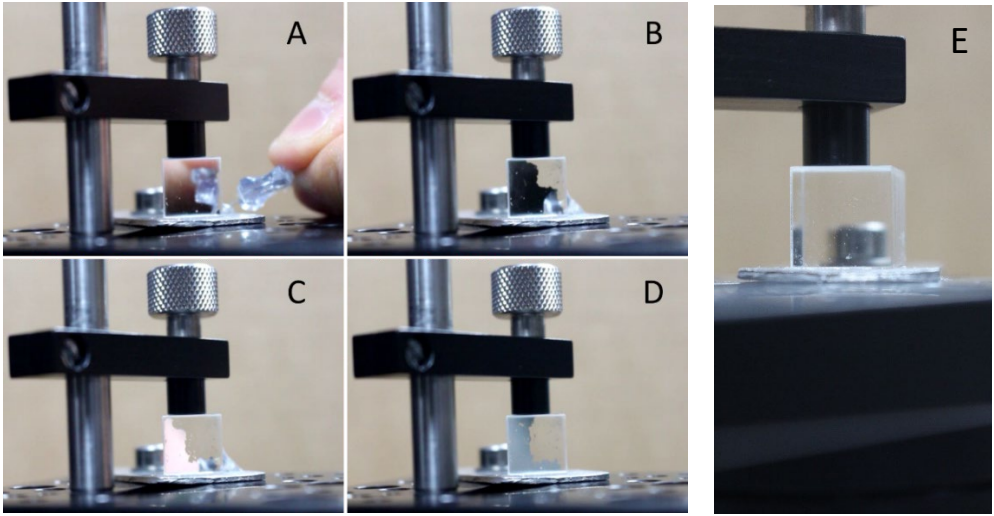

Figure 6. Subfigures A to C show the gradual infiltration of water between the prisms after leaving a drop of water next to their base. Subfigure D shows how water moves slightly back if the remaining water outside is removed. Subfigure E shows the prisms once the infiltration is complete.

For the study of transmittance as a function of the incident angle, the prisms were located on top of a rotating platform, which is also shown in Fig. 4. The incidence point of the laser beam on the interface between the two prisms must be aligned with the rotation axis, so that it does not change to a different point on the surface with a distinct associated thickness $d$. The chosen rotating platform is a Thorlabs CR1M-Z6, with 0.1° accuracy, which allowed to measure the external angles of incidence on the prism $\varphi_i$, that can be geometrically related to the internal angle of incidence $\theta_i$ (see Fig. 3). The relation between both angles depends on the prism apex angle $\alpha$, and is given by the following equation:

$$\theta_i = \alpha + \arcsin\left(\frac{n_{air} \sin \varphi_i}{n_p}\right) \quad (12)$$

This angle $\alpha$ of the first prism can be measured using several optical methods. In our case, it was obtained from the experimental reflectance and transmittance measurements together with the distance $d$ between the prisms, as will be explained below. The transmitted and reflected intensities were measured using a couple of Thorlabs S120-UV photodiodes. The powers measured by these devices ($P_R$, $P_T$) were operated as shown below to obtain the values of reflectance and transmittance for each incident angle:

$$T = \frac{P_T}{P_R + P_T} \quad (13)$$

We present as our experimental result the comparison between experimental transmittance measurements and the theoretical dependences for the best couple of ($\alpha$, $d$) values that could be found. Both graphs show transmittance as a function of the normalized propagation constant $b$, within a small interval around its critical value. As predicted from Eqs. (7) and (8), $T$ shows oscillations for incidence angles below $\theta_{cr}$ ($b > 1$) and drops rapidly towards zero as soon as they exceed that critical value ($b < 1$). These features can be observed in Fig. 7 and 8, which correspond to the cases of air and water occupying the thin gap between both prisms (low index medium).

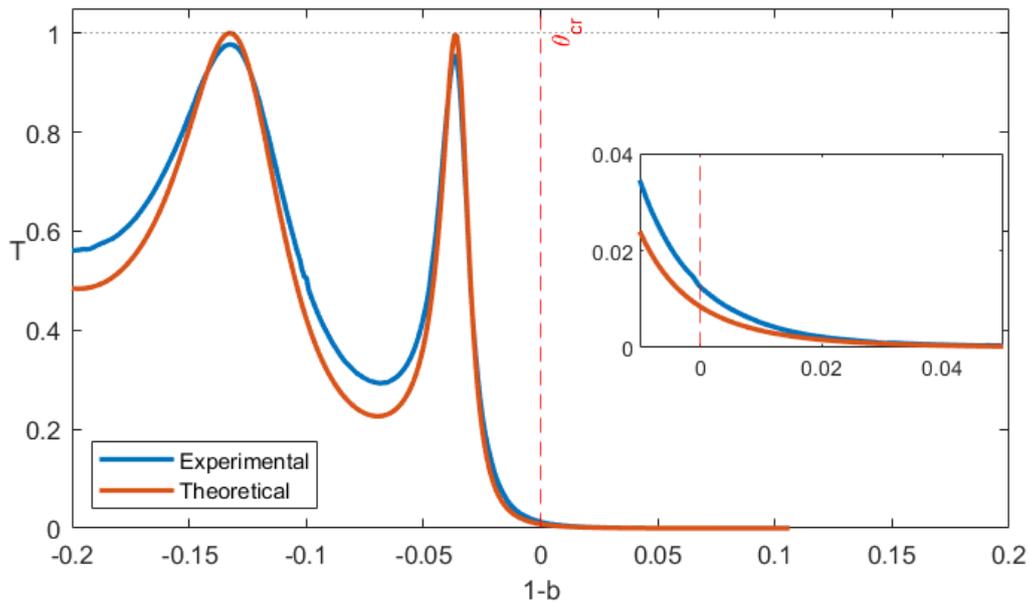

Figure 7. Experimental measurement for the transmittance of the system with an intermediate low index air layer compared to the most similar theoretical function found, corresponding to optimized parameters $\alpha = 44.9°$, $d = 1.56\,\mu m$.

To find the best values for both the prism angle $\alpha$ and the thickness of the low index layer $d$ we have followed a simple iterative procedure that can be perfectly performed by undergraduate students. Knowing that $\alpha \sim 45°$ for the isosceles right-angled prisms, we can choose that as an initial value, which we assume as true in order to obtain the internal angles for a first calculation of $d$. The calculated value for $d$ is the one minimizing the absolute distance between the experimental and theoretical positions of the transmittance maxima which are closer to the total internal reflection region. Once that optimum thickness is obtained, an analogue calculation is performed to optimize the choice of the prism angle. That new value would allow to repeat the previous calculation of the best possible $d$, repeating the procedure until the two values converge after a few iterations.

We remark that the resulting optimized prism angle values $\alpha$ are equal in both experiments with air and water ($\alpha = 44.9°$), within the precision of the rotation platform. The optimized values for the thickness of the intermediate layer are also in agreement with what was expected, because the prisms were separated and pressed against each other between the two measurements, and the two obtained thicknesses should be different but within the same order of magnitude ($d_{aire} = 1.56\,\mu m$, $d_{auga} = 3.42\,\mu m$).

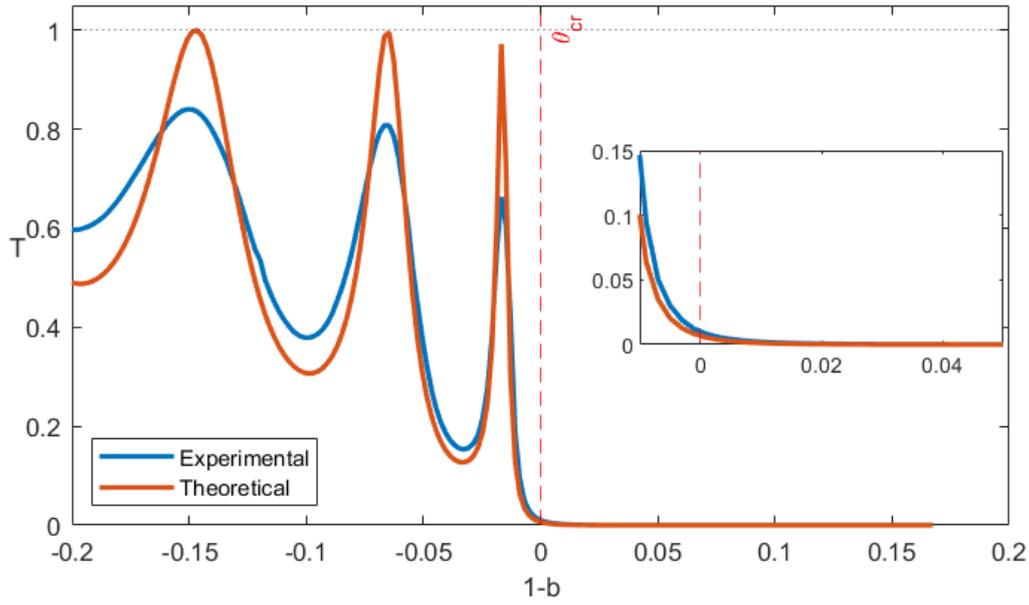

Figure 8. Experimental measurement for the transmittance of the system with an intermediate low index water layer compared to the most similar theoretical function found, corresponding to optimized parameters $\alpha = 44.9°$, $d = 3.42\ \mu m$.

Although the theoretical and experimental functions in Fig. 7 and 8 do not perfectly match, they are qualitatively similar enough to show the occurrence of the desired effects. However, there is an evident deviation between the two curves in both cases. There are several possible explanations for the observed differences, because of the limitations of our simple experimental device. First of all, the prism pair we used were found for a really affordable price, making them ideal for educational purposes, but also making the quality of their fabrication quite uncertain. Therefore, we are not sure about the flatness of their surfaces. Together with a non-exhaustive alignment of the rotation axis with the point of incidence at the internal interface of interest, this could have resulted in small random variations in the thickness of the intermediate low-index layer for the different external incident angles.

Moreover, our beam is a few millimeters wide, so it might be impinging on regions with different $d$ at the same time. This would produce a kind of blurring effect in the experimental maxima and minima, giving a possible reason for the experimental peaks to be wider. Besides, the beam may not be perfectly collimated, resulting in several incidence angles and producing a similar effect. This might also explain why experimental critical transmittance is higher than its theoretical value, since light incident at some angles slightly below $\theta_{cr}$.

## 4. Remarks on the analogy with quantum tunneling

The analogy between the optical behavior of electromagnetic waves when FTIR occurs at a three-layer system and the quantum behavior of a particle when facing a finite potential barrier has been widely studied using different approaches [8, 10, 11, 23]. The starting point is the isomorphism between the one-dimensional Helmholtz's equation for TE waves in Eq. (2) and the one-dimensional Schrödinger's equation for the wavefunction of a particle $\psi(x)$:

$$\frac{d^2\psi(x)}{dx^2} + \frac{2m}{\hbar^2}[E - V(x)]\,\psi(x) = 0 \tag{14}$$

where $m$ and $E$ stand for the mass and energy of the quantum particle, and $\hbar$ represents reduced Planck constant. The quantum potential $V(x)$ must correspond to a symmetrical potential barrier to fulfill the desired analogy, given by:

$$V(x) = \begin{cases} 0 & x \leq 0 \\ V_0 & 0 < x \leq d \\ 0 & x > d \end{cases} \tag{15}$$

The potential must be null at infinity to achieve a complete analogy between both problems. Thus, being able to strictly compare both Eqs. (2) and (14) requires subtracting in Eq. (2) the propagation constant at infinity, obtaining in the expression a term that vanishes for $x \to \infty$ just like $V(x)$ does. By doing so, the following equation is obtained:

$$\frac{d^2 E_y(x)}{dx^2} - [(\beta^2 - k_0^2 n_H^2) - (k_0^2 n^2(x) - k_0^2 n_H^2)]\, E_y(x) = 0 \tag{16}$$

Now, comparing Eqs. (10) and (12), some analogies between different magnitudes can be established:

$$\psi(x) \leftrightarrow E_y(x) \tag{17 a}$$

$$\frac{2m}{\hbar^2} V(x) \leftrightarrow -[n^2(x)k_0^2 - n_H^2 k_0^2] = -k_0^2[n^2(x) - n_H^2] \tag{17 b}$$

$$\frac{2m}{\hbar^2} E \leftrightarrow -[\beta^2 - n_H^2 k_0^2] = b\, k_0^2[n_H^2 - n_L^2] = k_0^2\, n_H^2 \cos^2\theta_i \tag{17 c}$$

It must be noticed that there is a relative sign between Eqs. (14) and (16) that has been considered when establishing the analogies. This sign is kept different in the equations due to the distinct typical conventions used in both fields of Optics and Quantum Mechanics. Eq. (17 b) means that an increase in the quantum potential is equivalent to a decrease in the refractive index. Similarly, Eq. (17 c) means that an increase in the energy of the particle is equivalent to an increase in the normalized propagation constant $b$, therefore corresponding to a decrease in the invariant $\beta$ and in the incidence angle $\theta_i$. Once these connections are established, studying the tunneling effect using FTIR becomes possible. According to this, while the usual $d$-scanning experiments on FTIR are analogous to varying the thickness of the quantum potential barrier, our experiment is equivalent to implementing variations in the energy of the quantum particle. Thus, the presented experiment provides the opportunity to further study the analogy between the two processes, which (as we have seen) is deeper and more interesting than it might seem at first glance.

Varying the energy of the incident particle keeping a constant potential barrier means a different kinetic energy, and therefore a change in velocity. Faster (and therefore more energetic) particles with energies $E < V_0$ are more likely to tunnel through the potential barrier. For particle energies above the height of the barrier, $E > V_0$ the transmission probability presents subsequent minima and reflectionless maxima, with the latter occurring at certain resonant energies.

## 5. Conclusions

A simple but slightly innovative experimental procedure for demonstrating and analyzing frustrated total internal reflection has been presented. An original interrogation method has been used, taking the incidence angle $\theta_i$ as the scanning variable, instead of the layer thickness $d$. Not taking the thickness $d$ as the interrogation variable allows us to eliminate the need for precise control of variations in the distance between the prisms. This method allows us to observe what happens for a constant $d$, not only for $\theta_i \geq \theta_{cr}$ but also for $\theta_i < \theta_{cr}$, when total reflection does not still occur. Consequently, we can see how the transmittance oscillates below the critical angle between subsequent maxima and minima of increasing contrast as $\theta_i$ approaches $\theta_{cr}$. Angle interrogation is quite interesting whether the analogy with quantum tunneling is considered, since it is analogue to an interrogation in quantum particle energy. Besides, normalized variables have been used, allowing us to ascertain more easily what the phenomenon of FTIR depends on. This has also shown the interesting possibility of quantifying the frustration of total internal reflection in terms of the value of the critical transmittance, which only depends on the normalized thickness $v$.

Finally, we would like to conclude by summarizing some conclusions drawn from the experimental results. In our experiment, different low index mediums between a couple of glass prisms have been used. We have been able to infiltrate water between the prisms using a simple method based on capillarity action. Interesting experimental curves have been obtained, in acceptable agreement with the expected theoretical variations. From these experimental measurements, the low-index layer thickness $d$ and the prism apex angle $\alpha$ have been obtained. Although the results seem qualitatively correct, and far enough for classroom demonstrations, they could be improved by a better alignment of the rotation axis with the incidence point at the interface of interest, or by ensuring the flatness of the prism surfaces (both conditions together guarantee that the thickness $d$ is kept constant while the incidence angle $\theta_i$ changes).

## Acknowledgements

C. F. R. would like to acknowledge the financial support of *Ministerio de Educación y Formación Profesional* through the grant 22CO1/012238. A. D. would also like to acknowledge the support of *Ministerio de Universidades* through FPU21/01302. Y. A. acknowledges a postdoctoral fellowship (ED481B-2021/027) from the Xunta de Galicia (Spain).